\documentclass[twocolumn,english,prb,showpacs]{revtex4}
\usepackage[T1]{fontenc}
\usepackage[latin9]{inputenc}
\usepackage{color}
\usepackage{amsmath}
\usepackage{graphicx}

\makeatletter

\providecommand{\tabularnewline}{\\}

\@ifundefined{definecolor}
 {\usepackage{color}}{}
\@ifundefined{definecolor}
 {\@ifundefined{definecolor}
 {\usepackage{color}}{}
}{}
\@ifundefined{definecolor}
 {\@ifundefined{definecolor}
 {\@ifundefined{definecolor}
 {\usepackage{color}}{}
}{}
}{}
\@ifundefined{definecolor}
 {\@ifundefined{definecolor}
 {\@ifundefined{definecolor}
 {\@ifundefined{definecolor}
 {\usepackage{color}}{}
}{}
}{}
}{}
\@ifundefined{definecolor}
 {\@ifundefined{definecolor}
 {\@ifundefined{definecolor}
 {\@ifundefined{definecolor}
 {\@ifundefined{definecolor}
 {\usepackage{color}}{}
}{}
}{}
}{}
}{}

\makeatother

\usepackage{babel}

\makeatother

\usepackage{babel}

\makeatother

\usepackage{babel}

\makeatother

\usepackage{babel}

\makeatother

\usepackage{babel}

\begin{document}

\title{Spin canting in a Dy-based Single-Chain Magnet \\
 with dominant next-nearest neighbor antiferromagnetic interactions}

\author{K. Bernot$^{1}$}

\author{J. Luzon$^{2}$}

\author{A. Caneschi$^{2}$}

\author{D. Gatteschi$^{2}$}

\author{R. Sessoli$^{2}$}

\email{roberta.sessoli@unifi.it}

\author{L. Bogani$^{3}$}

\author{A. Vindigni$^{4}$}

\author{A. Rettori$^{5}$}

\author{M. G. Pini$^{6}$}

\email{mariagloria.pini@fi.isc.cnr.it}

\affiliation{$^{1}$SCR/MI-INSA Laboratory, INSA-Rennes, CS 14315, 20 avenue des
Buttes de Coesmes, F-35043 Rennes, France \\
 $^{2}$Dipartimento di Chimica \& INSTM (UdR Firenze), Università
degli Studi di Firenze, I-50019 Sesto Fiorentino (FI), Italy}

\affiliation{$^{3}$Physikalisches Institut, Universität Stuttgart, Pfaffenwaldring
57, 70550 Stuttgart, Germany}

\affiliation{$^{4}$Laboratorium für Festkörperphysik, ETH Zürich, CH-8093 Zürich,
Switzerland}

\affiliation{$^{5}$Dipartimento di Fisica, Università degli Studi di Firenze,
I-50019 Sesto Fiorentino (FI), Italy}

\affiliation{$^{6}$Istituto dei Sistemi Complessi, Consiglio Nazionale delle
Ricerche, Via Madonna del Piano 10, I-50019 Sesto Fiorentino (FI),
Italy}

\date{\today}
\begin{abstract}
We investigate theoretically and experimentally the static magnetic
properties of single crystals of the molecular-based Single-Chain
Magnet (SCM) of formula {[}Dy(hfac)$_{3}$NIT(C$_{6}$H$_{4}$OPh)]$_{\infty}$
comprising alternating Dy$^{\text{3+}}$and organic radicals. The
magnetic molar susceptibility $\chi_{M}$ displays a strong angular
variation for sample rotations around two directions perpendicular
to the chain axis. A peculiar inversion between maxima and minima
in the angular dependence of $\chi_{M}$ occurs on increasing temperature.
Using information regarding the monomeric building block as well as
an \textit{ab initio} estimation of the magnetic anisotropy of the
Dy$^{\text{3+}}$ ion, this {}``anisotropy inversion'' phenomenon
can be assigned to weak one-dimensional ferromagnetism along the chain
axis. This indicates that antiferromagnetic next-nearest-neighbor
interactions between Dy$^{\text{3+}}$ ions dominate, despite the
large Dy-Dy separation, over the nearest-neighbor interactions between
the radicals and the Dy$^{\text{3+}}$ ions. Measurements of the field
dependence of the magnetization, both along and perpendicularly to
the chain, and of the angular dependence of $\chi_{M}$ in a strong
magnetic field confirm such an interpretation. Transfer matrix simulations
of the experimental measurements are performed using a classical one-dimensional
spin model with antiferromagnetic Heisenberg exchange interaction
and non-collinear uniaxial single-ion anisotropies favoring a canted
antiferromagnetic spin arrangement, with a net magnetic moment along
the chain axis. The fine agreement obtained with experimental data
provides estimates of the Hamiltonian parameters, essential for further
study of the dynamics of rare-earths based molecular chains.
\end{abstract}


\maketitle

\section{Introduction}

For many years, one-dimensional (1D) magnetic systems have been intensively
studied owing to their simplicity. A number of sophisticated theoretical
predictions have been experimentally verified as, for example, Haldane's
conjecture for Heisenberg antiferromagnets with integer spin \cite{Haldane},
Villain's mode for quasi-Ising $s=1/2$ antiferromagnets \cite{Villain_mode},
the contribution of solitons to thermodynamic properties for easy-plane
ferro- \cite{soliton_ferro} and antiferromagnetic \cite{soliton_antiferro}
systems with an in-plane symmetry-breaking magnetic field (or anisotropy).
Quasi-1D compounds, obtained by the molecular synthetic approach \cite{review_molecole,altra_revmolecole},
now allow the experimental investigation of dynamic phenomena that
had been theoretically predicted some decades ago \cite{Glauber}.
Such systems have been named Single-Chain Magnets (SCM's) \cite{rassegna,scm_clerac},
by analogy with Single-Molecule Magnets (SMM's) \cite{libro}, since
they show magnetic hysteresis with no evidence of 3D magnetic ordering
but rather due to slow dynamics of the magnetization of a pure 1D
character. It was shown \cite{scm_clerac,Cophome_EPL} that the relaxation
towards thermodynamic equilibrium in these systems is driven by thermally
activated single-spin flips, described in first approximation by the
stochastic theory developed by Glauber \cite{Glauber} for the ferromagnetic
1D Ising model. The effect of introducing non-magnetic impurities
in the chain was investigated \cite{Cophome_PRL,Cophome_PRB,Fsize_PRB},
too. The first system showing SCM behavior was the quasi-1D molecular
compound of formula {[}Co(hfac)$_{2}$NITPhOMe]$_{\infty}$,where
hfac is hexafluoroacetylacetonate, NITPhOMe is the nitronyl-nitroxide
radical 2-PhOMe-4,4,5,5-tetramethyl-4,5-dihydro-1H-imidazolyl-1-oxil-3-oxide\cite{Cophome}.
The repeating unit of such molecular chains is formed by two different
magnetic centers, a M$^{{\rm II}}$(hfac)$_{2}$ moiety containing
a metal (in the present case Co$^{\text{II}}$), and a radical moiety
that bridges the metals. The properties of this class of magnetic
polymers can then be tuned, with different effects, by rationally
changing either the metal or the radical moieties. In the rich field
that has stemmed from this first observation, several other compounds
presenting SCM behavior have been synthesized \cite{C1,C2,C3,C4,C5,C6,C7,C8,C9},
and evidences of quantum effects affecting the magnetization dynamics
at low temperatures have also been reported \cite{Clerac_PRL}. A
one-dimensional hysteretic behavior of the magnetization was identified
also in different non-molecular structures, such as atomic Co nanowires
of finite length, decorating the steps of vicinal Pt(997) surfaces
\cite{Gambardella}.

In general, to observe Glauber dynamics, two requirements are necessary:
(i) a strong Ising-like anisotropy and (ii) a very low ratio of interchain/intrachain
magnetic exchange interactions, in order for slow dynamics to be observed
above the transition temperature to 3D magnetic ordering. In the framework
of molecular engineering, quasi-1D chains in which the metal center
is a rare-earth (RE) ion are very appealing candidates to observe
a large variety of phenomena, as RE's are characterized by very different
magnetic anisotropies\cite{re_chains}. One limiting case is constituted
by molecular-based quasi-1D compounds of formula {[}(Gd)(hfac)$_{3}$NITR]$_{\infty}$
(where NITR is 2-R-4,4,5,5-tetramethyl-4,5-dihydro-1H-imidazolyl-1-oxil-3-oxide,
and R is ethyl (Et), isopropyl (iPr), methyl (Me), or phenyl (Ph)),
where the Gd(III) ions are magnetically isotropic. In these systems
helimagnetic behavior could be evidenced, due to competing magnetic
interactions between nearest neighboring (NN) and next-nearest neighboring
(NNN) spin sites. When R = Et, Villain's conjecture \cite{congettura}
of a two-step magnetic ordering to a low temperature helical 3D phase,
through an intermediate chiral spin liquid 3D phase, was confirmed
(the two transitions occurring at 1.88 K and 2.19 K, respectively)
\cite{PRL_congettura}. When passing from isotropic Gd to Dy, which
presents a strong Ising-like anisotropy, a transition to 3D magnetic
order occurring at 4.4 K can be observed in {[}Dy(hfac)$_{3}$NITEt]$_{\infty}$\cite{DyEt,DyNIT}.
However, using a bulkier NIT(C$_{6}$H$_{4}$OPh) radical \cite{DyPhOPh}
increases the distance between chains, so that SCM behaviour is observed
with no evidence of any 3D magnetic order down to the lowest investigated
temperature (1.7 K). The two requirements (i) and (ii), necessary
to observe Glauber dynamics, are thus fulfilled in {[}Dy(hfac)$_{3}$NIT(C$_{6}$H$_{4}$OPh)]$_{\infty}$.
Consistently, the relaxation time $\tau$ was found \cite{DyPhOPh}
to follow an Arrhenius law $\tau=\tau_{0}e^{\frac{{\Delta}}{{k_{B}T}}}$.
In order to fully understand the rich dynamics of these quasi-1D systems
\cite{DyPhOPh,family_lanthanides}, in particular the role played
by NNN interactions, a modelization of the static properties is necessary.
In this paper, an accurate experimental study of the angular dependence
of the single-crystal magnetization of {[}Dy(hfac)$_{3}$NIT(C$_{6}$H$_{4}$OPh)]$_{\infty}$,
combined with \textit{ab initio} estimation of the magnetic anisotropy
of the Dy$^{\text{3+}}$ ions, is presented. The static magnetic properties
of the chain system are simulated by means of a classical transfer
matrix calculation, formulated for a 1D Heisenberg model with antiferromagnetic
exchange coupling and non-collinear, local anisotropy axes. Fine agreement
with experimental data as a function of crystal orientation, temperature
and magnetic field is obtained and the Hamiltonian parameters can
then be estimated. The dominant role of NNN magnetic interactions
of these molecular structures, despite the large Dy-Dy separation,
is clearly demonstrated. These findings are expected to be extremely
important to target novel SCM dynamics in molecular compounds, and
to rationalize the SCM behavior of {[}Dy(hfac)$_{3}$NIT(C$_{6}$H$_{4}$OPh)]$_{\infty}$
and the related class of compounds \cite{family_lanthanides}. In
particular, as the role of natural and induced defects is substantially
different when competing NN and NNN interactions are present, this
situation can lead to the observation of different dynamic regimes
in addition to the already known ones.

\section{Crystal structure and \textit{ab initio} calculations}

Recently we investigated in detail a monomeric compound of formula
Dy(hfac)$_{3}$(NITR)$_{2}$ where the metal ion is coordinated to
the oxygen atoms of two different radicals\cite{monomero_R_Dy_R},
like in the chain. In fact, we suggest that the monomer can be considered
as a building block of the chain. Using angle-resolved magnetometry,
the molecular magnetic susceptibility tensor of the monomeric derivative
has been reconstructed, revealing an exceptionally large Ising-type
magnetic anisotropy of the Dy$^{\text{3+}}$ ion\cite{monomero_R_Dy_R}.
This technique also provides the orientation of the principal magnetic
axes of the monomer; the Ising axis turns out to lie in between the
two radicals, almost along the binary axis of the approximate c$_{2}$
symmetry of the molecule. Moreover, the experimental findings nicely
agree with \textit{ab initio} calculations within ca. 7\textdegree{}
deviation between the experimental and theoretical directions of the
magnetization easy-axis.

Here, we apply the same quantum chemistry method(CASSCF/RASSI-SO\cite{rassiso}),
using the MOLCAS-7.0 package\cite{molcas}, in order to obtain the
single-ion easy-anisotropy axes of the Dy$^{\text{3+}}$ ions in the
{[}Dy(hfac)$_{\text{3}}$NIT(C$_{6}$H$_{4}$OPh)]$_{\infty}$ chain
compound. In this multi-configurational approach relativistic effects
are treated in two steps, both based on the Douglas-Kroll Hamiltonian\cite{DKHamiltonian}.
Scalar terms are included in the basis set generation and are used
to determine spin-free wavefunctions and energies, through the use
of Complete Active Space Self Consistent Field (CASSCF) method. Then,
Spin-Orbit coupling is treated with a Restricted Active Space State
Interaction computation (RASSI-SO), which uses the CASSCF wavefunctions
as the basis states. By using the resulting eigenstates of the previous
method, the gyromagnetic tensor of the ground doublet Kramer's state
can be computed and diagonalized in order to obtain the three main
anisotropy axes and the gyromagnetic values along those axes (\textit{g$_{x}$},
\textit{g$_{y}$} and \textit{g$_{z}$}).

Computations are performed in a quantum cluster model considering
a Dy$^{\text{3+}}$ ion and its surrounding ligand molecules, i.e.
three hfac and two NIT radicals, using the geometry determined by
the X-ray structure analysis. In this cluster model several modifications
have been introduced in order to reduce the large computational time:
the fluorides in the hfac ligands have been replaced by H atoms, the
NIT(C$_{6}$H$_{4}$OPh) radicals have been replaced by NIT(C$_{6}$H$_{5}$),
and H atoms have been added to the external oxygen atoms of the radicals
transforming them in close-shell molecules, and therefore reducing
the active space of the CASSCF calculations. Despite the previous
modelization, the quantum cluster would still require a large computational
time due to its size. Therefore a Cholesky decomposition technique\cite{Cholesky},
recently implemented in the MOLCAS package, is used to overcome problems
arising from the large size of the two-electrons integral matrix.

All the atoms were represented by basis sets of atomic natural orbitals
from the ANO-RCC library as implemented in the MOLCAS-7.0 quantum
chemistry package and the following contractions were used: {[}8s7p4d3f2g]
for Dy, {[}3s2p] for O, C and N, and {[}2s] for H. The CASSCF active
space consisted on the Dy 4f orbitals, containing 9 electrons in 7
orbitals. The previous \textit{ab initio} computation in the monomeric
compound showed a minor role of the quadruplets and the doublets CASSCF
states in the Spin Orbit coupling effect on the energy levels of the
$^{\text{6}}$H$_{\text{15/2}}$ ground multiplet. Therefore, only
the three sextets, $^{\text{6}}$H, $^{\text{6}}$F and $^{\text{6}}$P,
have been computed in three different CASSCF state average calculations,
one for each of the sextets. The 21 resulting CASSCF states were introduced
in a RASSI-SO state interaction in order to compute the effect of
the Spin-Orbit coupling. The computed energies of the $^{\text{6}}$H$_{\text{15/2}}$
multiplet are listed in Table \ref{Elevels}.

\begin{table}[t]

\caption{\textit{Ab initio}-calculated energies of the eight doublets of the
$^{\text{6}}$H$_{\text{15/2}}$ multiplet of the Dy$^{\text{3+}}$
ions in the coordination environment determined from the crystallographic
structure.}

\begin{centering}
\begin{tabular}{|c|c|c|c|c|c|c|c|c|}
\hline
doublet  & E0  & E1  & E2  & E3  & E4  & E5  & E6  & E7\tabularnewline
\hline
\hline
Energy (K)  & 0  & 58.3  & 84.6  & 125.9  & 164.9  & 238.4  & 328.9  & 591.0\tabularnewline
\hline
\end{tabular}
\par\end{centering}

\label{Elevels}
\end{table}

\begin{figure}
\begin{centering}
\includegraphics[width=0.48\textwidth]{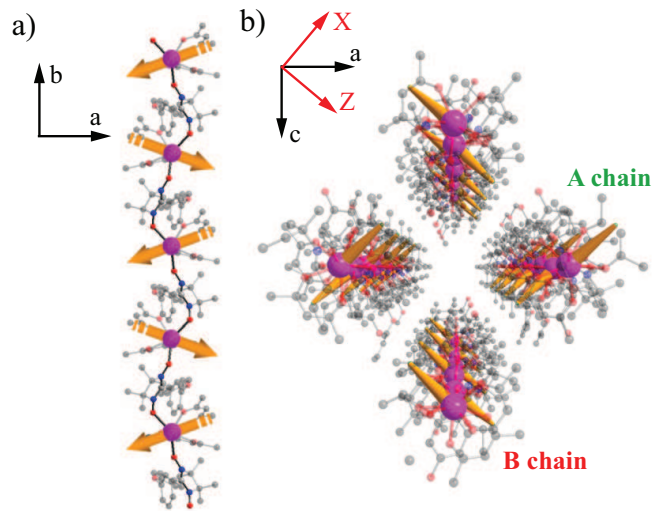}
\par\end{centering}

\caption{(color online) a) Representation of the Ising axes of the Dy$^{\text{3+}}$
ion, as gathered from \textit{ab initio} calculations and superimposed
on the chain in the $ab$ plane. b) View along the $b$ crystallographic
axis of the crystal packing, showing the two symmetry related types
of chains. The orientation of the $X$ and $Z$ axes in the \textit{ac}
plane is also represented (the $Y$ axis coincides with $b$). }

\label{fig:struct}
\end{figure}

The gyromagnetic factors for the ground doublet have also been evaluated
and found to be \textit{g$_{x}$}=0.3, \textit{g$_{y}$=}0.7, and
\textit{g$_{z}$}=18.7. The computed single-ion anisotropy axis almost
coincides with a binary axis in the idealized symmetry of the DyO$_{\text{8}}$
polyhedron, similarly to what observed for the monomer. The easy axis
forms an angle $\theta$ of ca. 80$^{{\rm o}}$
with respect to the $b$ crystallographic axis (along the chain axis),
whereas the projection of the single-ion anisotropy axis in the $ac$
plane (perpendicular to the chain axis) forms an angle $\phi$ of
ca. 50\textdegree{} with the $a$ axis. Compared to the triclinic
monomer, here the situation is complicated by the fact that the chain
crystallizes in the P$2_{1}2_{1}2_{1}$ space group (N$^{{\rm o}}$19),
Z= 4, with three two-fold screw axes as symmetry elements. The asymmetric
unit contains one Dy$^{\text{3+}}$ ion, and consequently all dysprosium
atoms are symmetry related. When superimposing the Ising axis of the
Dy$^{\text{3+}}$ ion on the chain, its symmetry elements generate
a canted structure as depicted in Fig. \ref{fig:struct}a. In Fig.
\ref{fig:struct}b, a view along the chain direction (the crystallographic
$b$ axis) of the crystal packing of the chains is reported. As a
consequence of this crystal packing, even if the chains are structurally
all equivalent, they are differently oriented from the magnetic point
of view. Two types of chains (${\rm A}$ and ${\rm B}$) are related
by a $2_{1}$ two-fold screw axis (in the $ac$ plane), while each
of these chains is generated by its own $2_{1}$ axis (along $b$).
When analyzing the magnetic behavior of the chain, we thus have to
consider not only the canted structure generated by the alternate
inclination of the Ising axes with respect to the chain direction,
$b$, but also the two families of chains (${\rm A}$ and ${\rm B}$),
whose projections on the $ac$ plane of the easy axis of the Dy$^{\text{3+}}$
ions form an angle of ca. 80\textdegree{}.

\section{Magnetic measurements}

Single-crystal magnetic measurements were performed using a homemade
horizontal rotator that allows, after orientation of the crystal,
to measure the magnetic susceptibility along and perpendicularly to
the chain axis. The morphology of the crystals, with large (101) and
($\overline{1}0\overline{1}$) faces, does not allow to perform rotations
along the principal crystallographic axes. We thus define a laboratory
frame, ($X,Y,Z$), with $Z$ corresponding to the normal to the (101)
crystal face, $Y$ = $b$, and $X$ orthogonal to the first two. The
new reference frame is thus obtained performing a rotation of the
crystallographic frame, ($abc$), around $b$ (the chain axis), by
an angle $\alpha\approx49.8^{{\rm o}}$ (see Fig. \ref{fig:struct}b).

\begin{figure}
\begin{centering}
\includegraphics[width=0.4\textwidth]{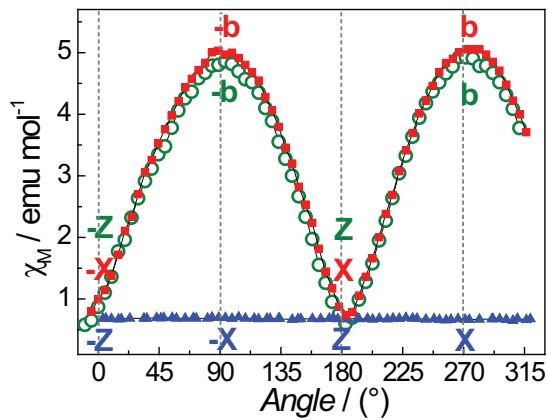}
\par\end{centering}

\caption{(color online) Angular variation of the molar susceptibility, $\chi_{M}$,
for three orthogonal rotations at 2.8 K, measured in a 1 kOe external
field. Rotations were performed around $X$ (green circles), $b$
(blue triangles) and $Z$ (red squares).}

\label{fig:magRotation}
\end{figure}

In Fig. \ref{fig:magRotation} we show the angular dependence of the
ratio $M/H$ between magnetization and field, from here on indicated
as the molar susceptibility $\chi_{M}$, in three orthogonal rotations
performed at 2.8 K in an external field of 1 kOe. As expected for
an orthorhombic system, the crystallographic axes correspond to relative
extrema. Two, out of three rotations, are almost identical, and display
a strong angular dependence of $\chi_{M}$, which has its maximum
along the chain direction. The third rotation, the one around $b$,
shows a very low, angle-independent value of $\chi_{M}$.

\begin{figure}
\begin{centering}
\includegraphics[width=0.45\textwidth]{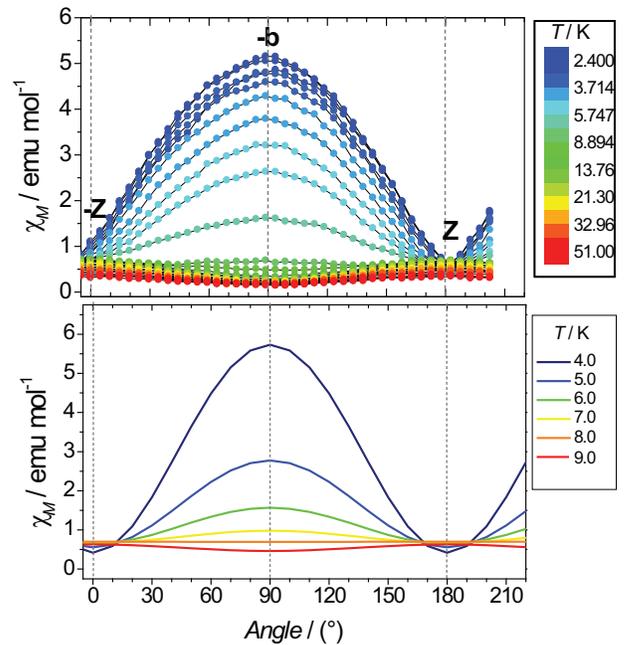} 

\par\end{centering}

\caption{(color online) (Top) Angular variation of the molar susceptibility,
$\chi_{M}=M/H$, for temperature $T$ ranging from 2.4 K (blue) to
51 K (red), measured in an external magnetic field of 1 kOe. Rotations
were performed around the $X$ axis, with $0^{{\rm o}}$ and $180^{{\rm o}}$
corresponding to the field aligned perpendicularly to the chain along
\textit{-Z} and \textit{Z}, respectively, whereas, at $90^{{\rm o}}$,
$\chi_{M}$ is measured along the chain. Temperature color mapping
is described on the right part of the figure. Vertical dot lines are
guides to the eyes. (Bottom) Transfer matrix simulation of the angle
dependence of $\chi_{M}=M/H$ (with $H=1$ kOe) when rotating around
$X$. The classical spin Hamiltonian, Eq. (\ref{Ham}), and the Hamiltonian
parameters specified in Section IV were used for the calculation.
Lines of different colors refer to different temperatures.}

\label{fig:ChiAngTemp}
\end{figure}

\begin{figure}
\begin{centering}
\includegraphics[width=0.45\textwidth]{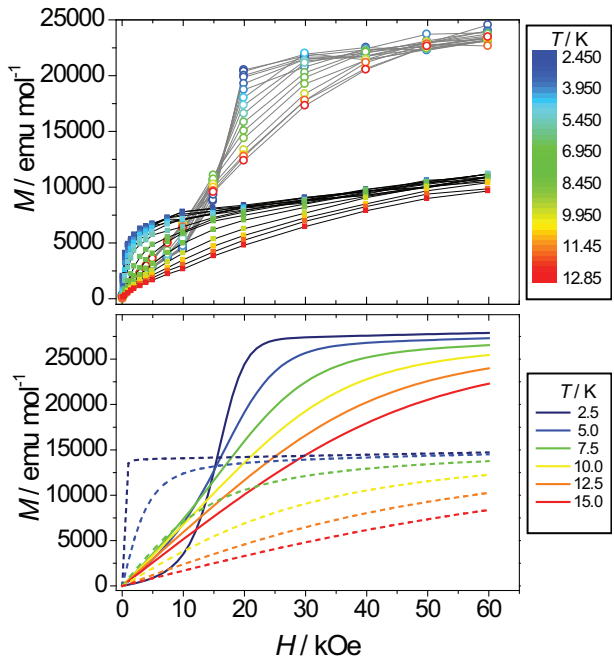} 

\par\end{centering}

\caption{(color online) (Top) Field dependence of the magnetization along the
chain (full squares) and perpendicular to the chain (along the $Z$
axis, open circles), for temperature \textit{T} ranging from 2.45
K (blue) to 12.85 K (red). The temperature color mapping is depicted
on the right. (Bottom) Transfer matrix simulation of the field dependence
of $M$ $vs$ $H$ along the chain direction (dashed lines) and perpendicular
to it (solid lines) obtained using the classical spin Hamiltonian,
Eq. (\ref{Ham}), and the Hamiltonian parameters specified in Section
IV. Lines of different colors refer to different temperatures.}

\label{fig:magTemp}
\end{figure}

The temperature dependent magnetic behavior of {[}Dy(hfac)$_{3}$NIT(C$_{6}$H$_{4}$OPh)]$_{\infty}$
was also investigated. Measurements were performed in the 2.4-51 K
temperature range, both for the angular variation of $\chi_{M}$ (Fig.
\ref{fig:ChiAngTemp}) and for the field ($H$) dependence of the
magnetization $M$ (Fig. \ref{fig:magTemp}). If we consider the rotation
around \textit{X} reported in Fig. \ref{fig:ChiAngTemp}(top) for
temperatures between 2.4 and 6.5 K, the maximum of $\chi_{M}$ is
found at $90^{{\rm o}}$, thus along the chain, and the minima at
$0^{{\rm o}}$ and $180^{{\rm o}}$, i.e. perpendicularly to the chain.
At 8 K, the $\chi_{M}$ curve has almost no angular dependence. For
temperatures from 9.5 K to 51 K, the maxima and minima are inverted
with respect to the low temperature curves. This is a quite spectacular
and peculiar feature which, at first glance, seems troubling. However,
previous investigations\cite{monomero_R_Dy_R,Mnchain} give us some
hints to rationalize this result:
\begin{itemize}
\item The direction of the Ising anisotropy of the Dy$^{\text{3+}}$ ion
in the monomer \cite{monomero_R_Dy_R} is temperature independent
in the temperature range investigated here, thus the anisotropy crossover
observed around 8 K cannot be attributed solely to an electronic effect
of the Dy$^{\text{3+}}$ ions in the chain compound.
\item The projection of the Ising axes, gathered from \textit{ab initio}
calculations and the analysis of the monomeric building block \cite{monomero_R_Dy_R}
of the chain structure suggests that the easy axes are not collinear.
\item The observed behavior (i.e., the anisotropy inversion) recalls the
one encountered in {[}{[}Mn(TPP)O$_{2}$PPh]$\cdot$H$_{2}$O]$_{\infty}$
(where TPP=$meso$-tetraphenylporphyrin), a Mn$^{{\rm III}}$-based
canted antiferromagnetic SCM \cite{Mnchain}. 

\end{itemize}
It is interesting to note that similar inversions (at two different
temperatures) were observed \cite{Giapponesi} in the magnetic torque
curves of TPP{[}Fe(Pc)(CN)$_{2}$]$_{2}$, a molecular conductor that
can be described by a one dimensional anisotropic Heisenberg model
with antiferromagnetic exchange interactions.

A scenario of a canted AF 1D structure similar to the one observed
in the {[}Mn(TPP)O$_{2}$PPh]$\cdot$H$_{2}$O chain \cite{Mnchain}
requires that, in the alternating Dy-radical chain under study, the
NNN AF interactions dominate the NN ones. To confirm this hypothesis,
the field dependence of the magnetization was measured. In Fig. \ref{fig:magTemp}(top)
we report the $M~vs~H$ curves recorded along and perpendicularly
to the chain. A very different shape of the curve is observed depending
on whether the field is aligned along the non-compensated moment (i.e.,
along the chain), thus giving a rapid saturation of the magnetization,
or closer to the compensated easy axes (i.e., perpendicularly to the
chain), providing a sort of metamagnetic transition when the magnetic
field overwhelms the antiferromagnetic interaction (see Section \ref{Theory and discussion}
for quantitative details).

The rapid saturation of the component along the chain observed at
3 K, with a plateau corresponding to ca 10000 emu mol$^{-1}$ ($\approx1.8\mu_{B}$),
is the typical behavior expected for a weak ferromagnet (WF) - indeed
in this case a 1D WF - along the direction of the non-compensated
moments. Perpendicularly to the chain, and more precisely along the
$Z$ axis, the magnetization is almost linear at low field, but shows
a rapid increase around 15 kOe and saturation at larger fields, almost
22000 emu mol$^{-1}$ ($\approx4.1\mu_{B}$).

Interestingly, the step in the magnetization perpendicular to the
chain becomes smoother when temperature is increased, and finally
disappears for temperatures above 10 K, when the weak AF interaction
becomes negligible and the anisotropy of the sample is driven by the
single-ion anisotropy of the Dy$^{\text{3+}}$. When the field is
applied along the chain direction, i.e. the direction of the non-compensated
moment, a more regular behavior is observed, as the initial susceptibility
monotonously decreases with increasing temperatures.

This feature sheds some light on the quite unique behavior of the
whole family of lanthanide-based SCM's
\cite{DyPhOPh,family_lanthanides}. Magnetization curves recorded
on the 4$f$-based SCM family always display steps at relatively
high field, and a trend was observed along the lanthanide series
\cite{DyPhOPh,family_lanthanides}. In fact, the position of the
step is explained by the simultaneous presence of an anisotropic
center (providing the anisotropic canted structure) and of an AF
interaction along the chain. Hence, the stronger is the intrachain
magnetic interaction, the higher is the field needed to compensate
this interaction, shifting the step towards higher fields.

This allows concluding that, in the {[}Dy(hfac)$_{3}$NIT(C$_{6}$H$_{4}$OPh)]$_{\infty}$
SCM under study:
\begin{itemize}
\item Antiferromagnetic interaction between Dy$^{\text{3+}}$ ions is present.
Due to this fact, the magnetization components perpendicular to the
chain direction cancel out at low temperature and low field, while
they become favored at higher temperatures.
\item The high temperature hard axis becomes the easy axis at low temperature
because non-compensation of the canted spins occurs.
\item The anisotropy inversion is visible around 9 K, suggesting a relatively
weak intrachain exchange interactions, as expected in compounds involving
tripositive lanthanides.
\end{itemize}
If we now compare the observed magnetization values for the chain
and the monomer derivative, a first saturation $M_{{\rm sat}}\approx58000$
emu mol$^{-1}$ ($\approx11\mu_{B}$) has been observed along the
easy-axis of the monomer \cite{monomero_R_Dy_R}. At a first level
of approximation we can neglect the radical contribution, which is
much smaller than that given by the paramagnetic ions; moreover, the
{}``radical'' sublattice is on itself experiencing AF interactions
mediated by the lanthanide ion and is also frustrated due to the sizeable
AF NNN interaction between Dy$^{\text{3+}}$ spins (see Section \ref{Theory and discussion})\cite{monomero_R_Dy_R}.
This allows providing a rough estimation of the angle, $\theta$,
formed by the easy axis of each Dy$^{\text{3+}}$ center with the
chain axis, $b$. In fact, comparing the afore-mentioned saturation
value ($\approx58000$ emu mol$^{-1}$) with the rapid saturation
one ($\approx$ 8000 emu$\cdot$mol$^{-1}$) reached when the field
is applied along the chain, yields $\theta=\cos^{-1}(8000/58000)\approx82^{{\rm o}}$.
This value is consistent with the results of the \textit{ab initio}
calculations.

\begin{figure}
\begin{centering}
\includegraphics[width=0.4\textwidth]{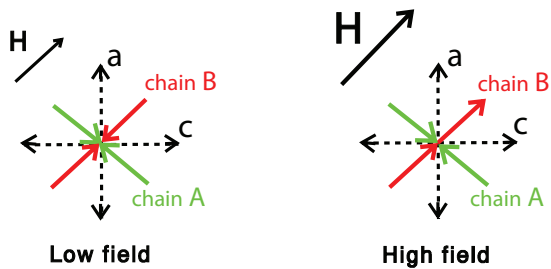}
\par\end{centering}

\caption{(color online) Influence of an applied magnetic field in the $ac$
plane on the orientation of the magnetic moments of neighboring Dy
ions of the two types of chains (A and B) .}

\label{fig:axisScheme}
\end{figure}

As for the magnetization curve measured perpendicularly to the
chain, the field was applied aligned along $Z$. Considering the
direction of the local easy axes predicted by the \textit{ab
initio} calculations (see Fig. \ref{fig:struct}b), the applied
field is almost parallel to the projection of these axes on the
$ac$ plane for type B chains, whereas it is almost perpendicular
to type A chains. For the two symmetry-related families of chains,
A and B, the projections of the local easy axes on the $ac$ plane
are almost reciprocally perpendicular but, remarkably, this is a
fully accidental coincidence rather than a symmetry-imposed
result. Therefore, as schematically represented in Fig.
\ref{fig:axisScheme}, the applied field can compete with the AF
interaction inside the chain of type ${\rm B}$ whereas it has no
effect on the other family (${\rm A}$), where Dy$^{\text{3+}}$
spins remain antiferromagnetically coupled. Thus, only one chain
type is contributing to the magnetization, which should approach a
value $M=(\frac{1}{2})58000\sin\theta$ (that is, $M\approx28700$
emu$\cdot$mol$^{-1}$), not very far from the experimental value.
Despite the fact that these figures can be only taken as
indicative, given the rough approximation of neglecting the
radical contribution, such saturation values for the magnetization
are in agreement with our prediction about the orientation of the
local easy axes, based on the \textit{ab initio} calculations.

\begin{figure}
\begin{centering}
\includegraphics[width=0.45\textwidth]{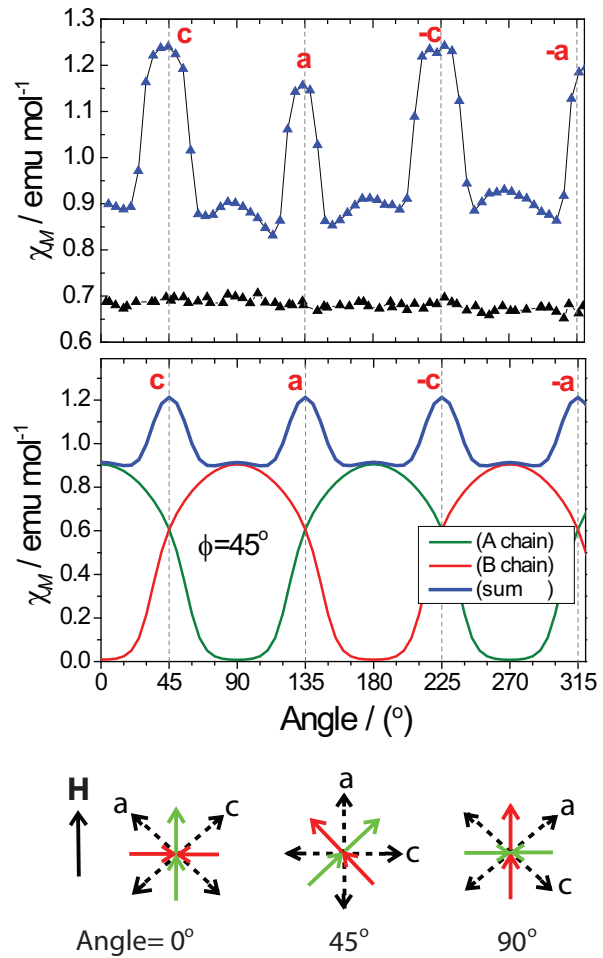} 

\par\end{centering}

\caption{(color online)(Top) Angular variation of the molar susceptibility
($\chi_{M}=M/H$) for the rotations around the chain axis $b$ at
2.5 K, measured in a 30 kOe external field (blue triangles) or in
a 1 kOe external field (black triangles). (Center) Transfer matrix
simulation of $M/H~vs~Angle$ at 2.5 K in a field of 30 kOe, showing
the contribution from each chain type (${\rm A}$ and ${\rm B}$),
as well as their sum. The classical spin Hamiltonian, Eq. (\ref{Ham}),
and the Hamiltonian parameters specified in Section IV were used for
the calculation. Perfect orthogonality 2$\phi$=$90^{{\rm o}}$ between
the projections on the \textit{ac} plane of the easy axes of the two
chain families $A$ and ${\rm B}$ was assumed. (Bottom) Schematic
representation of the corresponding spin configurations. }

\label{fig:acplane}
\end{figure}

A further confirmation of the correctness of this hypothesis can be
afforded by performing, at low temperature, a rotation around $b$
(the chain axis) under an external field of 30 kOe: i.e., a field
strong enough to be capable of distinguishing the different projections
of the Ising axes in the $ac$ plane by overcoming the AF interaction.
From Fig. \ref{fig:acplane}(top) one can clearly see that, contrarily
to what observed at low field, $\chi_{M}$ is now angle dependent,
with a very peculiar behavior with large maxima and a roughly $90^{{\rm o}}$
periodicity. It should be noticed that:
\begin{itemize}
\item The absolute maxima are observed along the crystallographic axes $a$
and $c$, and not along the directions where the projections of the
local axes fall.
\item The data show a secondary periodical structure of smaller maxima shifted
by an angle of ca. $45^{{\rm o}}$.
\end{itemize}
As a final remark, we notice that the maxima observed in Fig. 6(top)
along $a$ and $c$ are not strictly equivalent, but the direction
\textit{c} appears to be favored. In the following Section we will
show that this feature can be interpreted as due to a slight (but
appreciable) departure from orthogonality of the projection on the
$ac$ plane of the local easy axis of Dy$^{\text{3+}}$ ions belonging
to the two symmetry-related chain families, ${\rm A}$ and ${\rm B}$,
as indeed predicted by \textit{ab initio} calculations.

\begin{figure}
\begin{centering}
\includegraphics[width=0.45\textwidth]{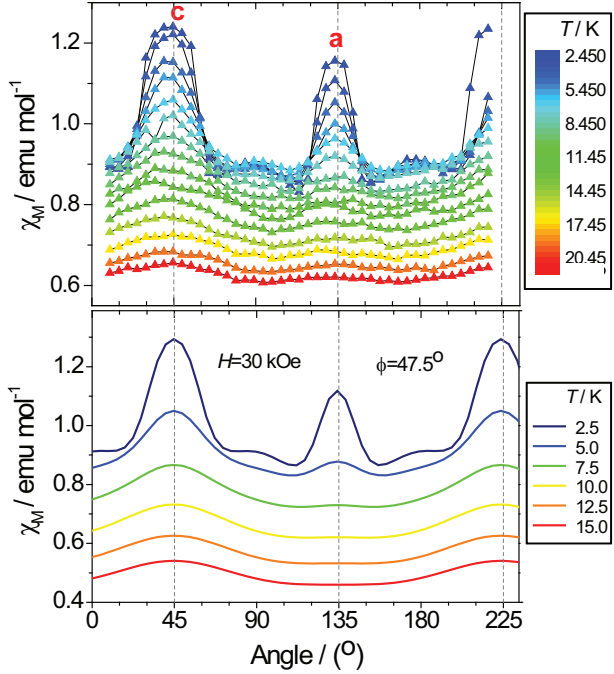} 

\par\end{centering}

\caption{(color online) (Top) Angular variation of the susceptibility $\chi_{M}=M/H$
from 2.45 K (blue) to 20.45 K (red) measured in a 30 kOe external
field. Rotations were performed around the chain axis $b$ as defined
in the text. Color mapping is described on the right part of the figure.
Solid lines are guides to the eyes. (Bottom) Transfer matrix simulation
of $\chi_{M}=M/H~vs~Angle$, at selected temperatures in a field of
30 kOe. The classical spin Hamiltonian, Eq. (\ref{Ham}), and the
Hamiltonian parameters specified in Section IV were used for the calculation.
Notice that, with respect to Fig. \ref{fig:acplane}(center), a slight
deviation from perfect orthogonality between the two chain families
was assumed, thus accounting for asymmetry in the height of the major
peaks. }

\label{fig:rotTemp}
\end{figure}

We have also checked the temperature dependence of the anisotropy
when rotating at high field around the chain axis, $b$ (see Fig.
\ref{fig:rotTemp}, top). The results show that the behavior is strongly
temperature dependent. Since in this temperature range the Dy$^{\text{3+}}$
anisotropy of the monomeric constituent units of the chain does not
change, these data confirm the key role played by the exchange interaction,
suggesting an order of magnitude of the AF interaction of a few Kelvins
(see the following Section for a more detailed discussion).

\section{Theory and discussion}

\label{Theory and discussion}

In order to interpret the measurements of the static magnetic properties
of the chain, we model both A and B chains with a classical spin Hamiltonian
\begin{eqnarray}
{\cal H} & = & -\sum_{i}\{J{\bf S}(i)\cdot{\bf S}(i+1)+D[S_{z_{i}^{\prime}}(i)]^{2}\nonumber \\
 & + & \mu_{0}\sum_{\alpha,\beta}H_{\alpha}g_{\alpha\beta}S_{\beta}(i)\}\label{Ham}\end{eqnarray}
 where $\mu_{0}$ is the Bohr magneton and $\alpha,\beta=X,Y,Z$ denote
Cartesian components in the crystallographic frame ($a$,$b$ and
$c$), while the subscript $z_{i}^{\prime}$ denotes the spin component
along a local anisotropy axis, varying with lattice site $i$. In
the sum, only the Dy centers are considered for the reasons given
below. The AF exchange interaction ($J<0$) is considered isotropic,
while the anisotropy is introduced through a single-ion term ($D>0$)
favoring local ${\bf z}_{i}^{\prime}$ axes, which are different for
even and odd sites, and in the Landé tensor ($g_{\alpha\beta}$).
In a local reference frame, the Landé tensor is assumed to have only
diagonal components, $g_{\Vert}$ and $g_{\perp}$ (where parallel
and perpendicular refer to the local axis ${\bf z}_{i}^{\prime}$).
Since the spin Hamiltonian (\ref{Ham}) is classical, the spin modulus
$S$ does not matter (its effect can be taken into account by the
parameters $J$, $D$, and $g$'s). For an easy comparison with the
calculation performed for the monomer \cite{monomero_R_Dy_R}, we
will assume $S=1/2$. Notice that the classical nature of the spins
allows the inclusion of single-ion anisotropy even in this case \cite{monomero_R_Dy_R}.
Concerning the neglect of the radicals, we note that their anisotropy
is very low compared to the Dy$^{\text{3+}}$ one, and the Dy-radical
interaction is necessarily frustrated, since the interaction between
the Dy$^{\text{3+}}$ ions is antiferromagnetic. As a first approximation,
the system can thus be considered as the superposition of two independent
one-dimensional magnetic lattices, one formed by the Dy$^{\text{3+}}$
ions (with magnetic anisotropy) and the other by the radicals (magnetically
isotropic). At low temperatures, the contribution of the {}``radical''
lattice is expected to be very small because of the relatively strong
NNN antiferromagnetic interaction between radicals \cite{monomero_R_Dy_R}.

Using a transfer matrix method \cite{Pandit}, the static magnetic
properties of the 1D classical spin model (\ref{Ham}) can be exactly
expressed in terms of the eigenvalues and eigenvectors of a real,
positive-definite matrix (see Appendix A for details). In this way,
starting from certain values of the Hamiltonian parameters, we were
able to numerically simulate the static magnetic behavior of the real
chain compound. The simulation procedure started from placing the
local easy axes at $\theta=75^{{\rm o}}$ from the $b$ axis and its
projection on the \textit{ac} plane at $\phi=\pm45^{{\rm o}}$ from
$a$. The other Hamiltonian parameters used for the simulation were
$J=-24$ K, $D=160$ K, $g_{_{\Vert}}=21$ and $g_{\perp}=4$, where
parallel and perpendicular refer to the local $z_{i}^{\prime}$ axis.
The contributions from the two families of chains were then summed
up. In Fig. \ref{fig:ChiAngTemp}(bottom) we report a simulation of
the quantity $M/H$, calculated for different temperatures when rotating
along $X$, to be compared with the measured $dc$ susceptibility
in Fig. \ref{fig:ChiAngTemp}(top). The susceptibility presents a
maximum along $b$ and a minimum along $Z$ for $T<8$ K, while the
trend is reversed for $T>8$ K. Thus, the anisotropy inversion - observed
at $T=8.5$ K on the experimental data in Fig. \ref{fig:ChiAngTemp}(top)
- is nicely reproduced.

The field dependence of the magnetization at selected temperatures,
both for $H$ parallel and perpendicular to the chain, was simulated
using the same Hamiltonian parameters: see Fig. \ref{fig:magTemp}(bottom).
The measured overall behavior of Fig. \ref{fig:magTemp}(top) was
reproduced, although a more quantitative comparison was prevented
by the neglect of the radical contribution in the adopted model (\ref{Ham}).

In particular, a relationship between the field at which a rapid increase
of the magnetization perpendicular to the chain is observed and the
Transfer-Matrix-computation parameters can be deduced by simple arguments.
Without loss of generality, we assume that spins are essentially constrained
to be oriented along their local anisotropy axes at low temperatures
and consider an applied field lying on the \textit{ac} plane at $\phi=+45^{{\rm o}}$
from $a$. This implies that the Zeeman energy of half of the spins
is frustrated (the ${\bf S}^{II}$-spin sublattice with the notation
of Appendix A). The flip of such spins lowers the Zeeman energy per
spin by a factor $\Delta E_{Zeeman}=2Sg_{_{\Vert}}\mu_{B}H\sin\theta$.
On the other hand, the corresponding increase of the exchange energy
per spin is $\Delta E_{exch}=4S^{2}\vert J\cos(2\theta)\vert$. The
reversal of such a sublattice is expected to occur when these two
energies are equal, i.e. at the field $H_{flip}=\frac{2S\vert J\cos(2\theta)\vert}{g_{_{\Vert}}\mu_{B}\sin\theta}=15.25$
kOe. These naïve energy considerations give the correct value for
the {}``jump'' field (see Fig. \ref{fig:magTemp}(bottom)), the
effect of temperature being solely that of rounding the step in the
magnetization curve.

Under quite general conditions \cite{Schrieffer}, the correlation
length $\xi$ of anisotropic, classical-spin chains behaves asymptotically
at low temperatures as $\xi\sim\exp\left(\frac{\Delta E_{exch}}{2k_{B}T}\right)$.
Consequently, for $T\ge\frac{\Delta E_{exch}}{2k_{B}}$, 
both families of spin chains are expected to lose their genuine 1D
character and respond to a field as independent, anisotropic paramagnetic
centers. Again, the order of magnitude of $\frac{\Delta E_{exch}}{2k_{B}}=10.4$
K matches with the temperature at which the anisotropy inversion was
observed.

The rotation around the chain axis, $b$, of the molar susceptibility
was also simulated in a strong external field ($H=30$ kOe), and the
contributions from each family of chains (${\rm A}$ and ${\rm B}$)
are reported in Fig. \ref{fig:acplane}(center), along with their
sum. A qualitative explanation of the observed behavior can be obtained
by looking in detail at the effect of the field when it is applied
at different angles in the \textit{ac} plane: see Fig. \ref{fig:acplane}(bottom).
When the field is along $a$ (or $c$), it lies at nearly $45^{{\rm o}}$
from the easy axis of both families of chains. Even if a smaller component
of the field (indeed proportional to $\cos45^{{\rm o}}$) acts on
the magnetic moments, this is still enough to overcome the AF interaction.
This is for example represented by the arrows at $Angle=45^{{\rm o}}$
in Fig. \ref{fig:acplane}(bottom). Given the large anisotropy of
Dy$^{\text{3+}}$ in this environment, the magnetic moments remain
aligned along their easy axis, and only the weak intrachain AF interaction
is overcome. The largest magnetization value we can measure is therefore
reduced to $M_{{\rm sat}}\sin75^{{\rm o}}\cos45^{{\rm o}}\approx39600$
emu mol$^{-1}$, not too far from the observed value ($\approx$ 37500
emu mol$^{-1}$). It is evident that each chain type gives a contribution
with a periodicity of 180$^{{\rm o}}$ so that, owing to the accidental
orthogonality between the two families ${\rm A}$ and ${\rm B}$,
the overall periodicity is $\approx90^{{\rm o}}$.

Finally, in order to simulate the asymmetry in the height of the major
peaks displayed in Fig. \ref{fig:acplane}(top), the hypothesis of
a perfect orthogonality between the projections of the two chain types,
${\rm A}$ and ${\rm B}$, was released. In Fig. \ref{fig:rotTemp}(bottom)
the angular variation of $\chi_{M}$ at selected temperatures, calculated
assuming an angle $\phi$ between $a$ and the projection of the local
easy axes in the $ac$ plane for the A-type chain slightly larger
than 45$^{{\rm o}}$, is reported. The fine agreement with experimental
data (see Fig. \ref{fig:rotTemp}, top) proves that the different
height of the major peaks at 45$^{o}$ and 135$^{o}$ can be attributed
to non-perfect orthogonality. Interestingly, even the minor maxima
between two main maxima are reproduced at sufficiently low temperatures.

\section{Conclusions}

In conclusion, we have performed a thorough experimental and theoretical
investigation to rationalize the complex magnetic behavior of the
Dy-based SCM {[}Dy(hfac)$_{3}$NIT(C$_{6}$H$_{4}$OPh)]$_{\infty}$.
By means of angle-resolved magnetometry the magnetic anisotropy of
this complex system has been accurately determined and it has been
possible to evidence a peculiar {}``anisotropy-inversion'' phenomenon,
in which the easy and hard axes of the magnetization of the system
swap when rising the temperature. These features could be fully explained,
with excellent agreement between theory and experiments, using a combined
theoretical approach in which the single RE-ion anisotropy is described
at the quantum-chemical level while the thermodynamic properties of
the whole system of coupled ions are computed with a classical spin
Hamiltonian. To our knowledge, this is the first report of such a
combined theoretical strategy, which we found indeed mandatory to
tackle the complexity of some magnetic molecular materials, often
characterized by a low symmetry environment of the magnetic centers.
The rich magnetic behavior of {[}Dy(hfac)$_{3}$NIT(C$_{6}$H$_{4}$OPh)]$_{\infty}$
was well reproduced by the model calculations that highlighted the
presence of two sets of non-interacting parallel chains, with their
projections mutually tilted by about 90$^{o}$. 
Accordingly, the metamagnetic transition inside each symmetry-related
family of chains could be addressed independently. The combined approach
has also allowed quantifying the sign, magnitude and role of NNN interactions
between Dy$^{3+}$ ions. Despite the large distances between the RE's
(> 8 Å), the overall behavior is actually dominated by this antiferromagnetic
NNN exchange pathway, which gives rise to the very rich magnetic behavior
observed. The same evidence was previously reported for isotropic
Gd ions. The present finding shows that NNN are more common than believed
so far and may induce a critical revision of the description of the
magnetic properties of polynuclear compounds. The present study may
thus represent a methodological guide line to rationalize the magnetic
properties of RE-based SCM's as well as other extremely complex behaviors
displayed by molecular materials in general. To the aim of understanding
the dynamic behavior of RE-based SCM's, the obtained information represents
a fundamental, preliminary step; from the knowledge that competing
NN and NNN interactions are present, for instance, one can reasonably
expect the observation of novel dynamic regimes. In particular, we
expect the effect of natural and induced defects to be substantially
different than the already known cases in which only NN interactions
are present\cite{Cophome_PRL,Cophome_PRB,Fsize_PRB}. As an example,
preliminary results indicate that, while the contribution of the radicals
to the static properties is almost negligible, they actively contribute
to the correlation length even when the Dy$^{3+}$ ions are partially
substituted by diamagnetic ions. In this sense, by doping with diamagnetic
ions the controlled interaction between segments of chains can be
tuned, thus allowing a further form of engineering of the dynamic
properties.

\acknowledgments

We acknowledge financial support from the NE-MAGMANET
(FP6-NMP3-CT-2005-515767) and the German DFG (SPP1137). J. L.
thanks the European Commission for the support through a Marie
Curie Intra-European Fellowship for Career Development.

\appendix*

\section{Transfer Matrix Method}

In this Appendix, we briefly describe the transfer matrix method,
which allows the determination, at any nonzero temperature, of the
static properties of a one-dimensional model of classical spins in
terms of the eigenvalues and eigenfunctions of a real, positive-definite
matrix. Here the method will be applied in the thermodynamic limit,
but it can be reformulated also to treat the case of a finite system
with open or periodic boundary conditions.

It is worth noticing that the method is exact: the only approximation
is that the spins are considered as classical vectors. Since the method
requires an integration over the unit sphere surface to be discretized,
numerical errors are significant only at very low temperatures, where
the integrand presents strong variations over the integration domain.
However, the accuracy of the method can be improved at will by using
a greater number of integration points.

Following Pandit and Tannous \cite{Pandit}, we consider a single
magnetic chain with \textit{two sublattices}, $I$ and $II$ (not
to be confused with the \textit{two families} of chains related by
{}``accidental'' orthogonality, ${\rm A}$ and ${\rm B}$).

\begin{figure}
\begin{centering}
\includegraphics[width=0.45\textwidth]{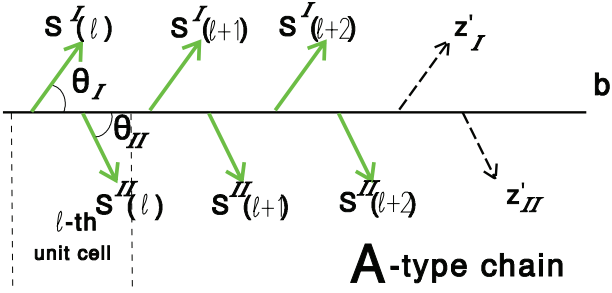}
\par\end{centering}

\caption{(color online) Canted ground state of the spins of a single chain
of type ${\rm A}$, described by Hamiltonian (\ref{Ham}). For each
unit cell $l$, there are two spins (thick, green arrows, representing
Dy$^{\text{3+}}$ ions) that belong to two different sublattices,
$I$ and $II$, with different local axes ${\bf z}_{I}^{\prime}$
and ${\bf z}_{II}^{\prime}$ (thin, dashed arrows).}

\label{fig:TM}
\end{figure}

To fix ideas, let us consider a single ${\rm A}$-type chain: in the
$l$-th unit cell, there are two spins, ${\bf S}^{I}(l)$ and ${\bf S}^{II}(l)$,
characterized by two different local axes, ${\bf z}_{I}^{\prime}$
and ${\bf z}_{II}^{\prime}$, forming equal angles ($\theta_{I}=\theta_{II}=75^{{\rm o}}$,
see Fig. \ref{fig:TM}) with the chain axis $b$ and their projections
in the $ac$ plane differing by $180^{{\rm o}}$ ($\phi_{I}=45^{{\rm o}}$,
$\phi_{II}=225^{{\rm o}}$) with the $a$ axis. %

The Hamiltonian of the chain can be rewritten, as in Ref. \onlinecite{Pandit},
in the form \begin{equation}
{\cal H}=\sum_{l=1}^{N}\{V_{1}[{\bf S}^{I}(l),{\bf S}^{II}(l)]+V_{2}[{\bf S}^{II}(l),{\bf S}^{I}(l+1)]\}\end{equation}
 where the total number of spins, in the single ${\rm A}$-type chain
considered, is $2N$ because there are two spins per each unit cell
($l=1,\cdots,N$) . In our case, $V_{1}$ and $V_{2}$ take the forms
($\alpha,\beta=X,Y,Z$)

\begin{eqnarray}
V_{1}[{\bf S}^{I}(l),{\bf S}^{II}(l)]=-J{\bf S}^{I}(l)\cdot{\bf S}^{II}(l)\nonumber \\
-\frac{D}{2}\{[S_{z_{I}^{\prime}}^{I}(l)]^{2}+[S_{z_{II}^{\prime}}^{II}(l)]^{2}\}\nonumber \\
-\frac{1}{2}\mu_{0}\sum_{\alpha,\beta}H_{\alpha}\{g_{\alpha\beta}^{I}S_{\beta}^{I}(l)+g_{\alpha\beta}^{II}S_{\beta}^{II}(l)\}\end{eqnarray}

\begin{eqnarray}
V_{2}[{\bf S}^{II}(l),{\bf S}^{I}(l+1)]=-J{\bf S}^{II}(l)\cdot{\bf S}^{I}(l+1)\nonumber \\
-\frac{D}{2}\{[S_{z_{II}^{\prime}}^{II}(l)]^{2}+[S_{z_{I}^{\prime}}^{I}(l+1)]^{2}\}\nonumber \\
-\frac{1}{2}\mu_{0}\sum_{\alpha,\beta}H_{\alpha}\{g_{\alpha\beta}^{II}S_{\beta}^{II}(l)+g_{\alpha\beta}^{I}S_{\beta}^{I}(l+1)\}\end{eqnarray}

Clearly, in the previous equations one has to take into account that
the crystallographic (laboratory) frame ($X,Y,Z$) and the local frame
($x^{\prime},y^{\prime},z^{\prime}$) are related by a rotation, so
that for the $I$-sublattice one has \begin{eqnarray}
S_{z_{I}^{\prime}}^{I} & = & \sin\theta_{I}\cos\phi_{I}S_{X}^{I}+\sin\theta_{I}\sin\phi_{I}S_{Y}^{I}\nonumber \\
 & + & \cos\theta_{I}S_{Z}^{I}\\
g_{XX}^{I} & = & g_{\perp}(\cos^{2}\theta_{I}\cos^{2}\phi_{I}+\sin^{2}\phi_{I})\nonumber \\
 & + & g_{\Vert}\sin^{2}\theta_{I}\cos^{2}\phi_{I}\\
g_{XY}^{I} & = & (g_{\Vert}-g_{\perp})\sin^{2}\theta_{I}\cos\phi_{I}\sin\phi_{I}\\
g_{XZ}^{I} & = & (g_{\Vert}-g_{\perp})\cos\theta_{I}\sin\theta_{I}\cos\phi_{I}\\
g_{YX}^{I} & = & g_{XY}^{I}\\
g_{YY}^{I} & = & g_{\perp}(\cos^{2}\theta_{I}\sin^{2}\phi_{I}+\cos^{2}\phi_{I})\nonumber \\
 & + & g_{\Vert}\sin^{2}\theta_{I}\sin^{2}\phi_{I}\\
g_{YZ}^{I} & = & (g_{\Vert}-g_{\perp})\cos\theta_{I}\sin\theta_{I}\sin\phi_{I}\\
g_{ZX}^{I} & = & g_{XZ}^{I}\\
g_{ZY}^{I} & = & g_{XZ}^{I}\\
g_{ZZ}^{I} & = & g_{\perp}\sin^{2}\theta_{I}+g_{\Vert}\cos^{2}\theta_{I}\end{eqnarray}
 and similar equations for the other ($II$) sublattice.

The partition function $Z={\rm Tr}~e^{-\beta{\cal H}}$ can be written
as \begin{eqnarray}
Z & = & \int d{\bf S}^{I}(1)\int d{\bf S}^{II}(1)\cdots\int d{\bf S}^{I}(N)\int d{\bf S}^{II}(N)\nonumber \\
 & \times & e^{-\beta V_{1}[{\bf S}^{I}(1),{\bf S}^{II}(1)]}e^{-\beta V_{2}[{\bf S}^{II}(1),{\bf S}^{I}(2)]}\cdots\nonumber \\
 & \times & e^{-\beta V_{1}[{\bf S}^{I}(N),{\bf S}^{II}(N)]}e^{-\beta V_{2}[{\bf S}^{II}(N),{\bf S}^{I}(1)]}\end{eqnarray}
 where periodic boundary conditions, ${\bf S}^{I}(N+1)={\bf S}^{I}(1)$,
were assumed \cite{Pandit}. A symmetric, positive-definite kernel
${\cal T}$ can now be defined 
\begin{eqnarray}
{\cal T}[{\bf S}^{I}(l),{\bf S}^{I}(l+1)]=\int d{\bf S}^{II}(l)\times\nonumber \\
e^{-\beta V_{1}[{\bf S}^{I}(l),{\bf S}^{II}(l)]}e^{-\beta V_{2}[{\bf S}^{II}(l),{\bf S}^{I}(l+1)]}\label{kernelA}\end{eqnarray}
 Denoting by $\lambda_{n}$ and $\Psi_{n}[{\bf S}^{I}(l)]$ the eigenvalues
and corresponding eigenfunctions of (\ref{kernelA}) \begin{eqnarray}
\int d{\bf S}^{I}(l+1) & {\cal T}[{\bf S}^{I}(l),{\bf S}^{I}(l+1)]~\Psi_{n}[{\bf S}^{I}(l+1)]\nonumber \\
= & \lambda_{n}\Psi_{n}[{\bf S}^{I}(l)]\label{eigenA}\end{eqnarray}
 one has that the largest eigenvalue $\lambda_{1}$ is always
non-degenerate, and the eigenfunctions satisfy the two conditions \begin{eqnarray}
{\cal T}[{\bf S}^{I}(l),{\bf S}^{I}(l+1)]=\sum_{n}\lambda_{n}\nonumber \\
\times\Psi_{n}^{*}[{\bf S}^{I}(l)]\Psi_{n}[{\bf S}^{I}(l+1)]\label{completeness}\end{eqnarray}
 \begin{equation}
\int d{\bf S}^{I}(l)\Psi_{n}^{*}[{\bf S}^{I}(l)]\Psi_{m}[{\bf S}^{I}(l)]=\delta_{n,m}\label{orthonormality}\end{equation}
 of completeness and orthonormality, respectively. Now, using first
(\ref{completeness}) to expand the kernel (\ref{kernelA}), and then
exploiting (\ref{orthonormality}), the partition function $Z$ can
be rewritten as \begin{align}
Z= & \int d{\bf S}^{I}(1)\int d{\bf S}^{I}(2)\cdots\int d{\bf S}^{I}(N)\times\nonumber \\
 & {\cal T}[{\bf S}^{I}(1),{\bf S}^{I}(2)]{\cal T}[{\bf S}^{I}(2),{\bf S}^{I}(3)]\cdots{\cal T}[{\bf S}^{I}(N),{\bf S}^{I}(1)]\nonumber \\
= & \sum_{n=1}^{\infty}(\lambda_{n})^{N}\to\lambda_{1}^{N}\end{align}
 where the largest eigenvalue $\lambda_{1}$ dominates in the thermodynamic
limit $N\to\infty$. Using a similar procedure, the crystallographic
components ($\alpha=X,Y,Z$) of the magnetization per site on the
$I$-sublattice can be expressed, for $N\to\infty$, as \begin{equation}
\langle S_{\alpha}^{I}\rangle=\int d{\bf S}^{I}\Psi_{1}^{*}({\bf S}^{I})S_{\alpha}^{I}\Psi_{1}({\bf S}^{I})\label{MA}\end{equation}
 In order to calculate the same quantities on the other ($II$) sublattice
of the single (${\rm A}$-type) chain considered, one defines another
symmetric, positive-definite kernel ${\cal U}$ and another integral
equation, respectively \begin{align}
 & {\cal U}[{\bf S}^{II}(l),{\bf S}^{II}(l+1)]=\int d{\bf S}^{I}(l+1)\nonumber \\
\times & e^{-\beta V_{2}[{\bf S}^{II}(l),{\bf S}^{I}(l+1)]}e^{-\beta V_{1}[{\bf S}^{I}(l+1),{\bf S}^{II}(l+1)]}\label{kernelB}\end{align}
 \begin{align}
\int d{\bf S}^{II}(l) & \Phi_{n}[{\bf S}^{II}(l)]~{\cal U}[{\bf S}^{II}(l),{\bf S}^{II}(l+1)]\nonumber \\
= & \lambda_{n}\Phi_{n}[{\bf S}^{II}(l+1)]\label{eigenB}\end{align}
 Notice that the eigenvalues of (\ref{eigenB}) and (\ref{eigenA})
are equal, so that the partition function $Z$ is the same, while
the eigenfunctions are different. Similarly to Eq. (\ref{MA}), the
crystallographic components ($\alpha=X,Y,Z$) of the magnetization
per site on the $II$-sublattice can be expressed, in the thermodynamic
limit, as \begin{equation}
\langle S_{\alpha}^{II}\rangle=\int d{\bf S}^{II}\Phi_{1}^{*}({\bf S}^{II})S_{\alpha}^{II}\Phi_{1}({\bf S}^{II})\label{MB}\end{equation}
 For small applied fields, the crystallographic components ($\alpha=X,Y,Z$)
of the molar susceptibility $\chi_{M}^{\alpha}$ for the single (${\rm A}$-type)
chain considered are then obtained from (\ref{MA}) and (\ref{MB})
as \begin{equation}
\chi_{M}^{\alpha}=\frac{1}{2}{N_{0}}\mu_{0}\frac{1}{{H^{\alpha}}}\sum_{\beta}\Big(g_{\alpha\beta}^{I}\langle S_{\beta}^{I}\rangle+g_{\alpha\beta}^{II}\langle S_{\beta}^{II}\rangle\Big)\end{equation}
 where $N_{0}$ is Avogadro's number.

\noindent The total susceptibility, to be compared with experimental
data, is then obtained by repeating the calculation for a chain belonging
to the other family \cite{notachaincalB}, i.e. ${\rm B}$ (related
to ${\rm A}$ by {}``accidental'' orthogonality), and then averaging
over the two contributions.

As regards the details of numerical calculations, following Pandit
and Tannous \cite{Pandit}, the integrals over the surface of a sphere
in (\ref{eigenA}) and (\ref{eigenB}) were approximated by using
McLaren's 72-points, 14th-degree formula. Thus, the eigenvalues and
eigenfunctions of the two integral equations (\ref{eigenA}) and (\ref{eigenB})
were obtained solving a (72$\times$72)-matrix eigenvalue problem
by standard numerical methods \cite{NumericalRecipes}. Because of
discretization, errors are significant only at very low temperatures,
where the integrands in (\ref{eigenA}) and (\ref{eigenB}) present
strong variations over the integration domains. However, the accuracy
of the method can be improved by using a greater number of integration
points.

\end{document}